\documentstyle[multicol,prl,aps,psfig,floats]{revtex}
\begin{document}
%\draft
\tightenlines
%\twocolumn[\hsize\textwidth\columnwidth\hsize\csname@twocolumnfalse\endcsname
\title{Coulomb Blockade and Insulator-to-Metal Quantum Phase Transition}
\author{Dmitri S. Golubev and Andrei D. Zaikin}
\address{Forschungszentrum Karlsruhe, Institut f\"ur Nanotechnologie,
76021 Karlsruhe, Germany\\
I.E.Tamm Department of Theoretical Physics, P.N.Lebedev
Physics Institute, Leninskii pr. 53, 117924 Moscow, Russia}

\maketitle

\begin{abstract}
We analyze an interplay between Coulomb blockade and quantum fluctuations
in a coherent conductor (with dimensionless conductance $g \gtrsim 1$)
attached to an Ohmic shunt. We demonstrate that at $T=0$ the system can be
either an insulator or a metal depending on whether its total resistance is
larger or smaller than  $h/e^2\approx 25.8$ k$\Omega$. In a metallic phase the
Coulomb gap is fully suppressed by quantum fluctuations.
We briefly discuss possible relation of this effect to recent experiments
indicating the presence of a metal-insulator phase transition in 2d disordered
systems. \end{abstract}


\begin{multicols}{2}
%]

%\textwidth=185mm \textheight=225mm \hoffset=-25mm \voffset=-10mm

It is well known that Coulomb interaction may strongly affect quantum transport
of electrons in mesoscopic conductors. For instance, electron tunneling across
metallic junctions can be strongly suppressed or even blocked completely at
$T=0$ due to Coulomb effects \cite{AL,SZ,GD}. This Coulomb blockade
of tunneling is a direct consequence of the electron charge discreteness,
manifestations of which persist even for junctions with resistances well below
the quantum resistance unit $R_{Q}=h/e^2\approx 25.8$ k$\Omega$. 

Recently it was argued \cite{GZ00} that the $I-V$ curve of an arbitrary
-- albeit relatively short -- coherent conductor in the presence
of interactions can be expressed in the form 
\begin{equation}
R\frac{dI}{dV}=1-\beta f(V,T),
\label{univ}
\end{equation}
where $1/R=(2e^2/h)\sum_nT_n$ is the Landauer conductance of a
scatterer and  $T_n$ are the transmissions of its conducting modes.
The magnitude of the interaction term in eq. (\ref{univ}) 
is controlled by the parameter 
\begin{equation} 
\beta=\frac{\sum_nT_n(1-T_n)}{\sum_nT_n},
\label{param}
\end{equation}
and $f(V,T)$ is a universal function which depends on $R$ as well as  
on the external impedance $Z_S(\omega )$ (e.g. leads) attached to the
scatterer. A similar result was also obtained in Ref. \cite{Alf} in the limit
of a single conducting mode, in which case $\beta =1-T_1$.

The parameter (\ref{param}) is already well known in the theory
of shot noise \cite{BB}. Since shot noise and interaction effects 
in mesoscopic conductors are both the manifestations of the electron charge
discreteness, it appears quite natural that the magnitude of both these
effects is governed by the same parameter $\beta$ (\ref{param}). 

It is also important that the above effects do not
disappear even if the charge quantization is not preserved in
a part of the system (e.g. in the leads). Consider an
important example of an Ohmic external impedance $Z_S(\omega ) \simeq R_S$ and
define $g_S=R_Q/R_S$ and $g_0=g+g_S$. The function $f(V,T)$ was already
evaluated before for the case of tunnel junctions. At not very high
energies ${\rm max}(T, eV) \ll g_0E_C$ and for $g_0 \gg 1$ one finds      
\begin{equation}
f(V,T)\simeq (2/g_0)\ln (g_0E_C/{\rm max}(T, eV)), 
\label{dif}
\end{equation} 
i.e. in this regime Coulomb interaction causes a logarithmic correction to
the scatterer conductance. This correction is small in the parameter
$1/g_0$. Here $E_C$ is the effective charging energy which sets the energy
scale for the interaction effects in our problem. The result (\ref{dif}) is
valid down to energies max$(T,eV) \sim g_0E_C \exp (-g_0/2)$. At even
lower $T$ and $eV$ energy relaxation effects turn out to be
particularly important making the conductance saturate
at a universal value \cite{GZ00} \begin{equation} 
G=\frac{1-\beta}{R}=\frac{2e^2}{h}\sum_nT_n^2, \label{lowT} \end{equation}
This formula recovers complete Coulomb blockade  in the limit $\beta \to 1$
(tunnel junctions), demonstrates the absence of it for ballistic scatterers 
($\beta \to 0$), and yields suppression of the Landauer conductance by the
factor 2/3 in the important case of diffusive conductors.

Within the approach \cite{GZ00} one can demonstrate that all corrections 
to eq. (\ref{lowT}) are small in the parameter $\beta /g_0 \ll 1$.
Nevertheless, one cannot yet conclude that the system
behavior is metallic at $T=0$ because the Langevin equation
analysis \cite{GZ00} does not fully account for the
charge discreteness in our system. Specifically, this analysis does not include
subtle nonperturbative effects which can be captured only by means
of the instanton technique. These effects may
provide a mechanism which turns a conductor into an insulator at $T=0$. Thus,
in order to understand the ground state properties of a system with disorder
and interactions, the instanton effects should be properly taken into account.

In this Letter we will analyze the role of the instanton effects
in relatively short coherent conductors. An important part of the
work was already carried out by Nazarov \cite{Naz} who
found the renormalized Coulomb gap for a coherent conductor in the form
$\tilde E_C \propto E_C\exp (-ag)$, where $a=\pi^2/8$ for diffusive conductors
\cite{Naz,Kam} and $a=1/2$ for tunnel junctions \cite{pa91}.
At temperatures below $\tilde E_C$ the charge quantization plays a crucial role
and should yield insulating behavior at $T=0$. 
We will extend the instanton analysis \cite{Naz} in several respects. Our main
goal is to study the effect of an external impedance $Z_S$ on the
renormalized Coulomb gap $\tilde E_C$. In particular, we will demonstrate that
for $R_S \leq R_Q$ quantum fluctuations yield {\it complete} suppression of
the Coulomb gap $\tilde E_C$ and, hence, destroy the insulating state even at
$T=0$.

{\it The model and effective action}. 
As in Ref \cite{GZ00} we will consider an arbitrary coherent scatterer between
two big reservoirs. Phase and energy relaxation are only allowed in the
reservoirs and not during scattering, i.e. 
the scatterer is assumed to be shorter than both dephasing and inelastic
relaxation lengths. Coulomb effects in
the scatterer region are described by an effective
capacitance $C$. The charging energy $E_C=e^2/2C$ is assumed to
be smaller than the typical inverse traversal time. 

The grand partition
function ${\cal Z}$ of the system ``scatterer+environment'' and the scatterer 
effective action $S_0$ can be written in the form \cite{SZ,SZ89,Z94}  
\end{multicols}
\begin{equation} {\cal Z}=\sum_{l=-\infty}^{\infty}\int_0^{2\pi l}{\cal
D}\varphi  \int {\cal D} q \exp \left[-\int d\tau 
\frac{C\dot\varphi^2}{2e^2}-S_0[\varphi ] +i\int d\tau\frac{\dot\varphi q}{e}  -
\frac{T}{2}\sum_{\omega }|\omega |Z_S(-i\omega )|q_{\omega }|^2\right],
\label{array1}
\end{equation}
\begin{equation}
S_0[\varphi ]=\sum_{m=1}^{\infty}\frac{D_mT^{2m}}{m}\int_0^{1/T}d\tau_1
\int_0^{1/T}d\tau_2...\int_0^{1/T}
d\tau_{2m}\frac{\sin [(\varphi (\tau_1)-\varphi (\tau_2))/2]}
{\sin [\pi T(\tau_1-\tau_2)]}...
\frac{\sin [(\varphi (\tau_{2m})-\varphi (\tau_1))/2]}
{\sin [\pi T(\tau_{2m}-\tau_1)]}.
\label{action}
\end{equation}

\begin{multicols}{2}
Here $\dot \varphi (\tau )/e = V(\tau)$ and $\dot q=I(\tau)$ are respectively
the fluctuating voltage across the scatterer and the current through the
system. Eq. (\ref{action}) was initially derived for
perfectly transparent contacts \cite{SZ89} and generalized in
Ref. \cite{Z94} to arbitrary transparencies. One has  
\begin{equation}
D_m=\frac{{\cal A}p_F^2}{2\pi}\int_0^1\theta d\theta D^m(\theta )=\sum_n (1-R_n)^m,
\label{Dm}
\end{equation}
where ${\cal A}$ is the cross section of the contact and $D (\theta )$ is its 
angle-dependent transparency. The sum $\sum_n$ is taken over independent
conducting channels and $R_n=1-T_n$.  

{\it Isolated scatterer}. Let us first consider an isolated scatterer 
$1/Z_S \to 0$ and rederive the result \cite{Naz} for $\tilde E_C$  
from our effective action (\ref{action}). In this limit integration over
$q$ in (\ref{array1}) fixes the charge to be constant 
$i\int d\tau(\dot\varphi q/e) \to 2\pi ilq_x/e$. At large 
conductances $g \gg 1$ the remaining integral over $\varphi$ is
evaluated within the saddle point approximation. We are interested in 
the nontrivial saddle points $\tilde \varphi (\tau )$ 
for the action $S_0$ which ``connect''
states with different winding numbers $l$. A general expression for such
instantons is given in \cite{Naz} and in a certain limit it reduces to a set
of Korshunov's instantons \cite{Kor} 
$\tilde \varphi (\tau ) = 2\arctan (\Omega (\tau -\tau_0))$. Since the
saddle point action $S_0[\tilde \varphi   ]$  does not depend on $\Omega$,
it suffices to set $\Omega \to 0$. Then $\tilde \varphi$ reduces further to a
set of ``straight lines'' $2\pi T\tau l$ \cite{FS}. Substituting $\tilde
\varphi (\tau ) =2\pi T\tau$ into (\ref{action}) one trivially gets
\begin{equation}
S_0[\tilde \varphi (\tau )]=\sum_{m=1}^{\infty}\frac{D_m}{m}=
-\sum_n\ln R_n.
\label{ln}
\end{equation}
Combining this equation with 
$\tilde E_C/E_C \propto \exp (-S_0 [\tilde \varphi (\tau )])$
one arrives at the result \cite{Naz} (for a spin degenerate case).

It is also possible to go beyond the exponential accuracy and to estimate the
pre-exponent in the expression for $\tilde E_C$. Here we restrict ourselves to
an approximation of non-interacting instantons. Within this approximation
$\tilde E_C$ can be found by means of a simple formula \cite{QPS} 
\begin{equation} 
\tilde E_C \sim T \left(\prod\limits_{k=1}^N
\frac{L_k}{z_{k}}\right) (S_0 [\tilde \varphi (\tau )]))^{N/2}\exp (-S_0 [\tilde \varphi (\tau )]),
\label{Ffinal} 
\end{equation} 
which yields correct results up to an unimportant numerical prefactor of order
one. Here $L_k$ and $z_{k}$ are respectively the effective volume and the
effective instanton size for the $k$-th zero mode and $N$ is the total number
of zero modes. Similarly to \cite{pa91} in our problem each instanton has two
zero modes, corresponding to shifts of its center $\tau_0$ in time from zero to
$1/T$ and to fluctuations of its frequency $\Omega$ within the interval $0\leq
\Omega \lesssim E_C$  (fluctuations with $\Omega > E_C$ are exponentially
suppressed by the charging term in the action). Thus we have
$N=2$, $L_1=1/T$ and $L_2 \sim E_C$.  The  parameters $z_{1,2}$ for both zero
modes can be easily evaluated. Substituting the eigenfunctions for these
zero modes $\partial \tilde \varphi /\partial \tau$ and $\partial \tilde
\varphi /\partial \Omega$ into the effective action one finds $z_1 \sim
1/\Omega$ and $z_2 \sim \Omega$. Thus, 
the  product $z_{1}z_{2}$ is $\Omega$-independent and just
reduces to a numerical factor of order one $z_{1}z_{2} \sim 1$. Then eq.
(\ref{Ffinal}) immediately yields      \begin{equation} 
\tilde E_C/E_C \sim \left[\prod_{n}R_n \right]
\ln \left[\prod_{n}R_n^{-1} \right]   \sim ag\exp (-ag).  
\label{rengap}
\end{equation} 
This formula is valid for $ag \gg 1$, i.e. either at large
conductances $g \gg 1$ or, if $g \sim 1$, for very small values $R_n$
implying $a \gg 1$. In a (spin-degenerate) single channel limit eq.
(\ref{rengap}) reduces  to $\tilde E_C/E_C \sim R_1 \ln R_1^{-1}$ in agreement
with the result derived in Ref. \cite{M} for $R_1 \ll 1$ within a
different technique. 

In order to find the dependence of the ground state energy $E_0$ on $q_x$ it is
necessary to sum over all possible instanton configurations.
For non-interacting instantons this summation yields 
\begin{equation}
E_0(q_x)=-\Delta \cos (2\pi q_x/e),  \;\;\;\; \Delta \sim \tilde E_C.
\label{E0}
\end{equation} 
In the particular case of a tunnel junction in the strong tunneling limit
$g \gg 1$ this result agrees with one derived in eq. (10) of Ref.
\cite{pa91}, where the pre-exponent in the expression for $\Delta$ 
(simply proportional to the inverse $RC$-time) was obtained
by means of an explicit calculation of the fluctuation determinants. We also
note that (comparatively weak) inter-instanton interaction
might slightly modify the result (\ref{E0}), both the form of $E_0(q_x)$ and
the pre-exponent in the expression for $\Delta$. However this effect will not
be important for us here.

Now we are in a position to study the scatterer conductance at
extremely low energies. In the limit of high external impedances
$Z_S \gg R_Q$ both the resistance $R(\omega )$ and conductance
$G (\omega )=1/R (\omega )$ of our scatterer are determined from the
correlation function $\langle \varphi \varphi \rangle$ evaluated for $1/Z_S
\to 0$. The calculation is again performed in imaginary time and is completely
straightforward. Similarly to Ref. \cite{ZP93} let us make a shift $q_x \to q_x
+\xi (t)$ in the exponent of eq. (\ref{array1}) and define the phase-phase
correlator 
\begin{equation}
\langle \varphi (\omega )\varphi (-\omega )\rangle = 
-\left(\frac{1}{\omega^2{\cal Z}} \frac{\delta^2 {\cal Z}}{\delta \xi_{\omega }
\delta \xi_{-\omega }}\right)_{\xi =0}.
\label{genfl}
\end{equation}
In order to evaluate the generating functional ${\cal
Z}[\xi (t)]$  (\ref{array1}) for $ag \gg 1$ it is
sufficient to consider small fluctuations of $\varphi$ and instantons.
Since we are mainly interested in the low frequency behavior of the
correlator (\ref{genfl}) we can safely assume that typical
frequencies of the source field $\xi_{\omega }$ do not exceed the
instanton ones $\Omega$. Under this adiabaticity condition our
previous instanton analysis can be trivially repeated.  Then for $q_x \to 0$ 
from eq. (\ref{genfl}) we obtain 
\begin{equation} 
\langle \varphi (\omega
)\varphi (-\omega )\rangle =\frac{e^2R_{eff}}{|\omega |} + \frac{4\pi^2\Delta
}{\omega^2}. \label{correl} \end{equation} 
Here $\omega$ is the Matsubara
frequency. What remains is to define the ``Matsubara resistance'' $\tilde R
(\omega )= (|\omega |/e^2)\langle \varphi (\omega )\varphi (-\omega )\rangle$
and perform an analytic continuation to real frequencies. After that at $T=0$
one gets 
\begin{equation}
G(\omega )=\frac{1}{R_{eff}+(2\pi /e)^2\Delta /i\omega }.
\label{gomega}
\end{equation}
This result implies that -- while at $\omega \gg \Delta$ the conductance
of our system remains finite -- in the opposite low frequency limit 
$\omega \ll \Delta$ the renormalized Coulomb gap $\Delta \sim \tilde E_C$
becomes important, the response is predominantly capacitive and $G(\omega )$
vanishes at $\omega \to 0$. At nonzero but low $T \ll \Delta$ and $1/Z_S\to
0$ one has $G(T) \propto \exp (-\Delta /T)$, i.e. the scatterer response
is {\it insulating} at $T=0$ due to Coulomb blockade.  
A simple estimate of the effective resistance $R_{eff}$ in
(\ref{correl}), (\ref{gomega}), similarly to \cite{ZP93}, would give 
$R_{eff}\simeq R$. In the spirit of our Langevin equation
analysis \cite{GZ00}, one can also expect a (frequency dependent) correction to
this estimate. In particular, at low $\omega < gE_C\exp (-g/2)$ it is natural
to expect $R_{eff}\simeq R/(1-\beta )$. 

{\it Metal-Insulator phase transition}.
Now let us see how the above behavior is modified in the presence of a finite
external impedance $Z_S(\omega )$. In this case the charge $q$ fluctuates
and, hence, should be treated as a quantum variable. For our purposes it is
convenient to first perform the path integral (\ref{array1}) over the phase
$\varphi (t)$. As before, in the limit $ag \gg 1$ we integrate over small
fluctuations of $\varphi$ and instantons. Integrating out small fluctuations,
for $|\omega |\lesssim E_C$ one arrives at an effective impedance  
\begin{equation} \tilde Z (-i\omega )\simeq Z_S (-i\omega )+ R,
\label{effimp} 
\end{equation} 
i.e. the scatterer impedance should simply be
added to one of the shunt. Evaluating the instanton contribution we
again assume that typical frequencies $\omega_q$
for the charge variable $q$ do not exceed $\Omega$. This is sufficient
for $Z_S \gg R$. Then, similarly to \cite{pa91}, we find
$$
{\cal Z}=\int{\cal D}q\exp \left(
-\frac{T}{2}\sum_{\omega }|\omega |\tilde Z(-i\omega )|q_{\omega
}|^2-\int d\tau E_0(q)\right). 
$$
Let us now choose $Z_S$ to be Ohmic $Z_S(\omega )\simeq R_S$. In this case the
resulting effective action for the charge $q$ coincides with one for a
linearly damped quantum particle in a periodic potential (\ref{E0}). This is a
well-known problem \cite{s,SZ} which can be treated, e.g., by means of the
standard renormalization group (RG) technique. Successively reducing the high
frequency cutoff $\omega_c$ and integrating out charges with higher $\omega_q$
one arrives at the RG equations \cite{s}  for $\tilde \Delta =\Delta /\omega_c$
and  $g_{\Sigma}=gg_S/g_0$:  \begin{equation} 
\frac{d\tilde \Delta }{d (\ln \omega_c)}=\tilde \Delta
(g_{\Sigma}-1)+O(\tilde \Delta^3 ), \;\;\; \frac{dg_{\Sigma}}{d(\ln
\omega_c)}=0. \label{ren}
\end{equation} 
Eqs. (\ref{ren})
demonstrate that for $g_{\Sigma}<1$ the value $\tilde \Delta$ -- though
initially small --  grows in the course of renormalization. Thus the Coulomb gap
remains finite and the system is an insulator at $T=0$. On the other hand, for
$g_{\Sigma}>1$ the Coulomb gap scales to zero. In this case 
Coulomb blockade is completely destroyed by quantum fluctuations and the
system behavior should be metallic down to $T=0$.

The charge-charge correlation function can also be derived in a straightforward
manner. At $|\omega | \lesssim E_C$ one finds
 \begin{equation} \langle qq \rangle_{\omega
}\simeq ((R+R_S)|\omega |+ \pi^2\Delta_r /e^2)^{-1}.
\label{qq}
\end{equation}
Here $\Delta_{r}$ plays the role of the renormalized Coulomb gap. It is equal
to $\Delta_{r} =\Delta  (\Delta /\omega_c)^{\frac{g_{\Sigma}}{1-g_{\Sigma}}}$
for $g_{\Sigma} < 1$ and $\Delta_r =0$ otherwise. As expected, charges are
localized in the insulating phase and delocalized in the metallic one.

Choosing the initial cutoff frequency as
$\omega_{c0} \sim E_C$, one can rewrite the first eq. (\ref{ren}) directly for the
combination $ag = \sum_n \ln R_n^{-1} $: 
\begin{equation} 
d (ag) /d (\ln \omega_c)=(1-g_{\Sigma})(1+1/ag).
\label{ren2} 
\end{equation}
This equation is valid as long as $ag \gg 1$. In the tunneling limit $T_n \ll
1$ one has  $\sum_n \ln R_n^{-1} \simeq
\sum_nT_n=g/2$, and (\ref{ren2}) reduces to the RG
equation for the tunneling conductance derived in \cite{pa91} as well as
in Refs. \cite{GS,HZ} (for the case $g_{\Sigma}=0$). 
%In other cases, however, 
%eq. (\ref{ren2}) determines only the scaling of
%$\sum_n \ln R_n^{-1} $ and not of the conductance $g=2\sum_nT_n$. 

Combining our present results with those of Ref. \cite{GZ00} we arrive 
at the following picture. In the
limit of large conductances $g_0 \gg 1$ the $I-V$ curve of an arbitrary
coherent scatterer is described by eqs. (\ref{univ})-(\ref{lowT}) down to
exponentially low temperatures and voltages max$(T,eV) \gtrsim \Delta_r$. In
this regime Coulomb effects lead to partial suppression of the scatterer
conductance which saturates at the value (\ref{lowT}) at low $T$ and $V$. For
$g_{\Sigma } \geq 1$ or if at least one of the channels is ballistic, $R_n=0$,
the Coulomb gap is fully suppressed $\Delta_r=0$, and the conductance remains
nonzero (\ref{lowT}) even at $T=0$. This is the metallic phase. On the other
hand, if $g_{\Sigma} <1$ and all $R_n >0$, at energies
$$
{\rm max}(T,eV) \lesssim \Delta_r \sim E_C [ag\exp
(-ag)]^{\frac{1}{1-g_{\Sigma}}} 
$$
Coulomb interaction leads to further suppression of the scatterer conductance
$G$ which eventually vanishes at $T=0$. This is the insulating phase. A
quantum phase transition between these two phases occurs either at
$g_{\Sigma}=1$ or if the conductance (per spin) of at least one of the channels
$g_n\equiv T_n$ becomes exactly equal to one.

Summarizing, an interplay between charge discreteness, coherent scattering and
Coulomb interaction yields two types of effects. One effect is controlled by
the parameter $\beta$ (\ref{param}) and results in partial suppression of the
system conductance at low $T$. For $g_0 \gg 1$ this suppression is fully
captured by the quasiclassical Langevin equation approach \cite{GZ00}. This
effect of the electron-electron interaction is relatively robust in the sense
that the interaction correction $\beta f$ (\ref{univ}) is never
suppressed unless {\it all} the channels are ballistic ($\beta =0$) and/or the
shunt conductance $g_S$ is infinite. Another effect is due to
instantons which give rise to the renormalized Coulomb gap  $\Delta_r$. This
interaction effect is much more subtle since the Coulomb gap can be destroyed
by quantum fluctuations much easier. It is this suppression which yields a
quantum  insulator-to-metal phase transition in our system.

Finally, let us briefly address possible implications of our
results for recent experiments \cite{Kr} which strongly indicate the presence
of a metal-insulator phase transition in various 2d disordered systems.
One can consider a (sufficiently small) coherent scatterer
with the dimensionless conductance $g$ viewing all other scatterers in the
system as an effective environment with the conductance $g_S$. On a
phenomenological level one can {\it assume} this environment to be Ohmic at
sufficiently low frequencies. {\it Under this assumption} one immediately
arrives at the conclusion about the presence of a quantum  metal-insulator
phase transition  at $g_{\Sigma}$=1. In 2d systems one has $g \sim g_S \sim
g_0 \sim g_{\Sigma }$. Therefore in such systems this phase transition should
be expected at conductances $\sim 1/R_Q$, exactly as it was
observed in numerous experiments \cite{Kr}. Local properties  
of the insulating and metallic phases are expected to be very different.
In the insulating phase charges should be localized around inhomogeneities
(puddles) due to Coulomb blockade, while in the metallic phase the Coulomb gap
is destroyed by quantum fluctuations and the charge distribution should be
much more uniform. These expectations are fully consistent with recent
experimental observations \cite{Ya}. Thus, there might be a direct relation
between the experimental results \cite{Kr,Ya} and the old problem \cite{s}
describing dissipative dynamics of a quantum particle in a periodic
potential. In order to explore this possibility a more detailed
theory, which would include, e.g., the issue of quantum decoherence, would be
warranted.

\end{multicols}

\end{document}